\documentclass[notitlepage,11pt,pra,tightenlines,unsortedaddress,nofootinbib,floatfix,showkeys]{revtex4-1}
\usepackage[unicode=true, pdfusetitle, bookmarks=true, bookmarksnumbered=false, bookmarksopen=false, breaklinks=true, pdfborder={0 0 0}, backref=false, colorlinks=true, linkcolor=blue, citecolor=blue, urlcolor=blue]{hyperref}
\usepackage{bm}
\usepackage{physics}
\usepackage{mathtools,mathrsfs,amssymb}
\usepackage{scalerel}
\usepackage{subdepth}
\setcitestyle{authoryear}
\DeclareMathAlphabet{\mathbbold}{U}{bbold}{m}{n}

\begin{document}

\title{Spin-aware movement of electrons and time-of-flight momentum spectroscopy}\thanks{For Andrea Oldofredi (ed.), \textit{Guiding Waves In Quantum Mechanics: 100 Years of de Broglie-Bohm Pilot-Wave Theory}, Oxford University Press, forthcoming}

\author{Siddhant Das}
\email{Siddhant.Das@physik.uni-muenchen.de}
\affiliation{Mathematisches Institut, Ludwig-Maximilians-Universit\"at M\"unchen, Theresienstr.\ 39, D-80333 M\"unchen, Germany}
\date{Jan.\ 24, 2023}

\begin{abstract}
In the framework of the de Broglie-Bohm pilot-wave theory, or Bohmian mechanics, we examine two pedagogical problems that illustrate the bound and unbound motion of spin-1/2 particles: First, a single spin-1/2 particle trapped in the ground state of a spherical box is studied in both the relativistic and nonrelativistic versions of the theory; second, the free time evolution of this particle once the confinement is released is examined, demonstrating how the Fourier transform of the prepared wave function yields the statistics of the particle's far-field (asymptotic) velocity, thereby providing a deeper understanding of time-of-flight momentum spectroscopy techniques.
\end{abstract}

\maketitle

\section{Introduction}\label{intro}
The de Broglie-Bohm (dBB) worldview is grounded in the mantra \emph{matter moves}. It is commonly expressed in terms of a dynamical theory for particles or fields---the elements of matter---and their quantum mechanical wave functions. The role of the wave functions is to steer matter in motion, analogous to, yet distinct from, the role played by electromagnetic or gravitational fields in classical mechanics. The most compelling, empirically successful version of the dBB (pilot-wave or Bohmian) theory features point-like particles pursuing continuous trajectories in spacetime, determined by their wave function; see \cite{Bohm1,Bohm2,BohmHiley,DurrTeufel,hollandbook,ShellyStanford,Bricmont,Oriols,struyvethesis}.

For a single particle with position \(\vb{R}\), the (Bohmian) trajectories solve the \emph{guiding equation} 
\begin{equation}\label{guide}
    \frac{\mathrm{d}}{\mathrm{d}\kern0.1em  t}\kern0.1em  \vb{R}(t) = \vb{v}^{\Psi}\big(\vb{R}(t),t\big)
\end{equation}
---a first-order equation of motion for the particle's velocity, where \(\vb{v}^{\Psi}(\vb{r},t)\) is a vector field that depends on the wave function \(\Psi\). The \(\Psi\) function satisfies Schr\"odinger's equation
\begin{equation}\label{sch}
    i\kern0.1em \hbar\,\frac{\partial\Psi}{\partial t} = \mathcal{H}\kern0.1em \Psi
\end{equation}
at \emph{all} times. (So in this theory ``wave-particle duality'' simply means that there is both a wave and a particle.) The type of wave function, the Hamiltonian \(\mathcal{H}\), and the velocity field \(\vb{v}^{\Psi}\), depend on the nature of particles, the presence of external potentials, and whether the motion is Galilei- or Lorentz-covariant. Their exact forms are educated guesses motivated by symmetry arguments, simplicity, empirical sufficiency, the anticipated classical limit, and arguably even operational input \cite{Wiseman2007}. Once specified, the dBB dynamics is determined by Eqs.\ \eqref{guide} and \eqref{sch}, and only these. The particle's trajectory \(\vb{R}(t)\) is thus uniquely defined for all \(t\), given suitable initial conditions \(\vb{R}(0)\) and \(\Psi(\vb{r},0)\).\footnote{\cite{Berndl,RodiTufel} establish the existence of the (\(N\)-body) dBB dynamics under very general assumptions on the external potentials and initial conditions.}

An example that is most commonly examined is a nonrelativistic spin-0 particle of mass \(m\) moving in an external potential \(V(\vb{r},t)\), for which \(\Psi\) is a complex-valued function of \(\vb{r}\) and \(t\),
\begin{equation}\label{velo0}
    \vb{v}^\Psi(\vb{r},t) = \frac{\hbar}{m}\kern0.1em \Im\kern-0.1em\left[\frac{\Psi^*\pmb{\nabla}\Psi}{\Psi^*\Psi}\right]
\end{equation}
or, alternatively, \(\smash{\vb{v}^\Psi(\vb{r},t) = \pmb{\nabla}S/m}\) writing \(\Psi = |\Psi|\exp(S/\hbar)\), and
\begin{equation}
    \mathcal{H} = -\,\frac{\hbar^2}{2\kern0.1em  m}\kern0.1em \nabla^2+V.
\end{equation}
The particle's motion is generally non-Newtonian, readily illustrated in the problem of quantum tunneling or barrier penetration; see \cite{Norsentunneling,Dewdneytunneling}. Now widely accessible, particle trajectories derived from the velocity field \eqref{velo0} offer an intelligible, paradox-free, and persuasive picture of various quantum phenomena. A notable recurring example is that of the double-slit experiment with (or without) monitoring; see \cite{Dewdney}, \cite{Laloe}.

Of course, the dBB theory does not stop at comforting pictures. As is widely recognized it yields the same experimental predictions as the textbook/standard quantum formalism \cite{Bohm2,DGZOperators}---thankfully, without glorifying the role of observation. An illustration of how this theory proceeds from pictures to predictions in the context of laboratory momentum measurements is provided below.

In this chapter, we will focus on the dBB equations for a single spin-1/2 particle, which are based on spinor-valued wave functions. Early on, it was thought that incorporating spin into the repertoire of the dBB theory would prove to be an impasse,\footnote{This may have been partially influenced by Bohm's presentation of Eq.\ \eqref{guide} in a second-order, Newtonian form \cite{Bohm1}, since it is generally regarded that spin is a type of ``nonclassical two-valuedness'' and a definitive symbol of the break from classical modes of thought; in part, this view may have also been reinforced by the ``no hidden variables'' theorem due to Kochen and Specker, which forbids assigning pre-existing values (such as the particle position \(\vb{R}\)) to all the spin components of a particle; see \cite{mermin}.} to wit: David Bohm reported in a letter, quoted in \cite[p.\ 97]{OlivalBohm}:
\begin{quote}
   ``von Neumann thinks my work correct, and even `elegant,' but he expects difficulties in extending it to spin.''
\end{quote}
But within a year, following his 1952 papers, Bohm demonstrated how the theory could naturally account for the behaviour of spin-1/2 particles \cite{BohmT}. Many physicists contributed to this subject in the years that followed, including Takehiko Takabayasi, Ralph Schiller, Jayme Tiomno, John S.\ Bell, and especially Peter R.\ Holland---it would be a worthy undertaking for historians to chronicle these developments.

Spin-dependent guiding equations have received relatively little attention in the dBB literature, outside of discussions of the EPRB-Bell or Stern-Gerlach experiments (see, e.g., \cite{Norsenspin,HollandNature}). Of greater concern, in almost all instances where spin-1/2 particles are discussed, the analysis is based on an inadequate nonrelativistic velocity field, which contradicts the relativistic velocity in the nonrelativistic limit, cf.\ \cite{Holland92,Holland,Hollanduniqueness}. This oversight, while usually not compromising much conceptual accuracy, is regrettable because atoms are composed entirely of spin-1/2 particles, which are most relevant to chemistry\footnote{See Lombardi \& Fortin's contribution in this volume.} and, for that matter, nearly everything. The subject surely deserves better treatment!

What better way is there of getting acquainted with the dBB dynamics of fermions than the simple setting of a single spin-1/2 particle trapped within a box (one of the first problems a student will encounter in quantum mechanics)? The particle-in-a-box problem is certainly of historical significance considering the interpretive debates between Born, Einstein, and Bohm, the specifics of which we will not address here (but see \cite[Sec.\ 3.1]{Wane}, \cite[Ch.\ XI, Sec.\ 3]{dBBook}, \cite[Sec.\ 6.5.3.]{hollandbook}).

The motion of a spin-1/2 particle (henceforth referred to as an electron) prepared in the ground state of a spherical box is investigated here in both nonrelativistic (Sec.\ \ref{NRT}) and relativistic (Sec.\ \ref{RT}) versions of the dBB theory. In both cases, the Bohmian trajectories are explicitly computable in a coordinate-free way---a rare luxury. The presentation assumes some familiarity with Pauli and Dirac's equations and is at a level of quantum mechanics and mathematical methods typical of courses taken by first-year graduate students in physics.

Switching gears, in Sec.\ \ref{deconfinement}, we consider the response of the electron to a sudden removal of the confining potential. While the Bohmian trajectories are no longer analytically solvable, it is shown that the escaping electron exhibits free-Newtonian behaviour in the far field---a hallmark of unbounded dBB motion in scattering situations \cite{DDGZ96,DDGZ}.

This problem offers a segue to examine a widely used technique for determining the momentum of an abruptly freed bound electron by measuring its arrival (or flight) time on a distant detector (Sec.\ \ref{momentumTOF}).\footnote{This is in contrast to the dBB treatments of momentum measurements discussed, e.g., in \cite[Sec.\ 6.8.]{hollandbook}, \cite[Sec.\ 15.1.2.]{DurrTeufel}, \cite[Sec.\ 2.A.3.]{Bricmont}, or \cite[Sec.\ 5.5.]{DGZOperators}, which involve a position measurement at a late but \emph{fixed time} rather than a time-of-flight measurement at a specified detection surface.\label{ToFconfuse}} Section \ref{anapplication} repurposes the electron-in-a-box example to illustrate this experimental technique, recovering the expected quantum mechanical result \emph{without reference to a quantum observable}. Here, we outline the typical steps involved in applying the dBB theory to experiments, explaining how seemingly random experimental outcomes can arise within a deterministic theory.

Our analysis highlights a crucial point: the dBB theory not only makes it possible to predict experimental results but also validates or limits the assumptions that experimenters invariably make in transforming their raw data (or the directly measured quantities) into final reported results. This is an important nuance that is often overlooked when comparing different ``interpretations'' of quantum mechanics.

Final concluding thoughts are offered in Sec.\ \ref{conclusion}.
\section{Electron-in-a-box: Nonrelativistic treatment}\label{NRT}
For an electron of mass \(m\) moving at nonrelativistic speeds, the wave function is a two-component spinor (or two-spinor) \(\smash{\Psi=\big(\psi_+(\vb{r},t),\psi_-(\vb{r},t)\big)^\top}\). The dBB velocity field and Hamiltonian are given, respectively, by \cite[Ch.\ 10]{BohmHiley}, \cite[Ch.\ 2]{struyvethesis}
\begin{equation}\label{velo}
    \vb{v}^\Psi(\vb{r},t)=\frac{\hbar}{m}\kern0.1em \Im\kern-0.1em\left[\kern-0.1em\frac{\Psi^\dagger\pmb{\nabla}\Psi}{\Psi^\dagger\Psi}\kern-0.1em\right]\,+\,\frac{\hbar}{2\kern0.1em  m}\kern0.1em \frac{\pmb{\nabla}\times\big(\Psi^\dagger\bm{\sigma}\Psi\big)}{\Psi^\dagger\Psi},
\end{equation}
and
\begin{equation}\label{Ham}
    \mathcal{H} = -\kern0.1em \frac{\hbar^2}{2m}\big(\bm{\sigma}\cdot\pmb{\nabla}\big)^2 + V(\vb{r},t)\kern0.1em \mathbbold{1}.
\end{equation}
Here, \(V\) is an external potential, \(\mathbbold{1}\) denotes the \(2\times2\) unit matrix (often omitted, but to be understood), and \(\bm{\sigma}=\sigma_x\vu{x}+\sigma_y\vu{y}+\sigma_z\vu{z}\) is Pauli's spin-vector matrix. Incorporating \eqref{Ham} into \eqref{sch} gives the Schr\"odinger-Pauli equation.\footnote{If a magnetic field \(\vb{B}\) is present, the gradient in Eqs.\ \eqref{velo} and \eqref{Ham} must be replaced by the gauge covariant derivative involving the vector potential (minimal coupling prescription). The Pauli interaction term \(\smash{\propto\,\bm{\sigma}\cdot \vb{B}}\) is automatically generated by this and need not be put in by hand.} The first (second) term on the right-hand side of \eqref{velo} is also called the \emph{convective} or \emph{osmotic} (\emph{spin} or \emph{Gordon}) velocity. The convective term is a natural generalization of the spin-0 dBB velocity \eqref{velo0} to spinor-valued wave functions.

\emph{A word of caution}: The spin velocity term is omitted in most discussions of particles with spin in the dBB literature. As emphasized in Sec.\ \ref{intro}, without it the resultant velocity field would not be recovered from the relativistic velocity field (Eq.\ \eqref{BD}, below) in the nonrelativistic limit. The latter is, in fact, \emph{uniquely pinned down} by covariance considerations \cite{Holland92,Holland,Hollanduniqueness}, as is, therefore, its nonrelativistic limit given by \eqref{velo}.\footnote{One can arrive at \eqref{velo}, or equivalently the current density \(\vb{j}=(\Psi^\dagger\Psi)\kern0.1em \vb{v}^\Psi\), using the nonrelativistic Schr\"odinger-Pauli theory; see \cite{Mike}.}

In dropping the spin term (or by inadvertently using the spin-0 equations), one faces the unexpected consequence that an electron in the hydrogen ground state is stationary! This mistake has practically become lore, inspiring novel proposals for modifying the dBB guiding equations to cure the ``stationary electron malady'', e.g., \cite{finley}. Nevertheless, as demonstrated in \cite{Colijn,Colijn1,Colijn2} (see also footnote \ref{omega}), if the full velocity field \eqref{velo} is taken into account the electron is in fact \emph{not} stationary in the hydrogen ground state.

We turn now to a discussion of an electron trapped in an infinite spherical well (or spherical box) of radius \(a\). For this, assume
\begin{equation}\label{sphericalwell}
    V(r) = \begin{cases}\,0, &r\le a\\\,\infty, &r>a\end{cases},\qquad r=\norm{\vb{r}},
\end{equation}
in \eqref{Ham}; the Hamiltonian simplifies to
\begin{equation}
    \mathcal{H} = -\,\frac{\hbar^2}{2m}\nabla^2 + V(r),
\end{equation}
in view of the identity \(\smash{(\bm{\sigma}\cdot\pmb{\nabla})^2=\nabla^2\mathbbold{1}}\)---a special case of the spin-matrix identity
\begin{equation}\label{identity}
    (\vb{a}\cdot\bm{\sigma})\kern0.1em (\vb{b}\cdot\bm{\sigma})=(\vb{a}\cdot\vb{b})\kern0.1em \mathbbold{1}+i\kern0.1em (\vb{a}\times\vb{b})\cdot\bm{\sigma}.
\end{equation}
The Schr\"odinger-Pauli equation thus reduces to two uncoupled Schr\"odinger equations for the spinor components \(\psi_\pm\), allowing for simple \emph{space-spin-separated} or \emph{spin-polarized} two-spinor solutions of the form
\begin{equation}\label{gs}
    \Psi(\vb{r},t) = e^{-\,itE/\hbar}\,\psi(\vb{r})\chi,
\end{equation}
where \(\psi\) is a solution of the time-independent Schr\"odinger equation \(\mathcal{H}\psi = E \psi\), and \(\chi\) is an otherwise arbitrary, constant two-spinor. (\(\psi\) is commonly referred to as the coordinate wave function or the ``spatial-part'' of \(\Psi\), while \(\chi\) is called the ``spin-part'' of \(\Psi\).) The ground state wave function, given by \cite[p.\ 130]{Grif}, is distinguished by its simple dependence on \(r\) (in particular, it exhibits no zeros or nodes within the spherical box): 
\begin{equation}\label{psit}
    \psi(\vb{r}) = \frac{1}{\sqrt{2\kern0.1em \pi a}~ r}\kern0.1em \sin(\frac{\pi\kern0.1em  r}{a})
\end{equation}
for \(\smash{r<a}\), and zero otherwise. It corresponds to
\begin{equation}\label{energy}
    E = \frac{\hbar^2\pi^2}{2\kern0.1em  m\kern0.1em  a^2}.
\end{equation}

For spin-polarized wave functions, the guiding equation simplifies to
\begin{align}\label{guide1}
    \frac{\mathrm{d}}{\mathrm{d}\kern0.1em  t}\kern0.1em  \vb{R}(t)=\frac{\hbar}{m}\kern0.1em \Im\!\left[\frac{\pmb{\nabla}\psi}{\psi}\right]\!\big(\vb{R}(t),t\big)\,+\,\frac{\hbar}{2\kern0.1em  m}\left[\frac{\pmb{\nabla}\times\left(|\psi|^2\kern0.1em \vu{s}\right)}{|\psi|^2}\right]\!\big(\vb{R}(t),t\big),
\end{align}
introducing the constant \emph{unit} vector
\begin{equation}\label{spinvec}
    \vu{s} = \frac{\chi^\dagger\bm{\sigma}\chi}{\chi^\dagger\chi}
\end{equation}
---the so-called unit spin-vector associated with \(\Psi\).\footnote{For any two-spinor \(\smash{\chi=(\chi_+,\chi_-)^\top}\), \(\smash{\chi^\dagger \bm{\sigma} \chi = 2\kern0.1em \Re\big[\chi_+^*\chi_-\big]\kern0.1em \vu{x}+2\kern0.1em \Im\big[\chi_+^*\chi_-\big]}\kern0.1em \vu{y}+\big(|\chi_+|^2-|\chi_-|^2\big)\kern0.1em \vu{z}\); consequently 
\(
    \norm*{\chi^\dagger\bm{\sigma}\chi}^2= 4\kern0.1em |\chi_+^*\chi_-|^2 +\big(|\chi_+|^2-|\chi_-|^2\big)^2=|\chi_+|^4+2\kern0.1em  |\chi_+|^2\kern0.1em  |\chi_-|^2+|\chi_-|^4 = \big(\chi^\dagger\chi\big)^2 \Rightarrow \norm{\vu{s}}=1.
\)} The numerator of the last term in Eq.\ \eqref{guide1} is equivalent to \(\big(\kern0.1em \pmb{\nabla}|\psi|^2\big)\times\vu{s}\), since \(\smash{\pmb{\nabla}\times\vu{s}=\vb{0}}\) (the components of \(\vu{s}\) do not depend on the spatial coordinates). We thus have
\begin{align}\label{guide2}
    \frac{\mathrm{d}}{\mathrm{d}\kern0.1em  t}\kern0.1em  \vb{R}(t)=\frac{\hbar}{m}\kern0.1em \Im\!\left[\frac{\pmb{\nabla}\psi}{\psi}\right]\!\big(\vb{R}(t),t\big)+\frac{\hbar}{m}\kern0.1em \Re\!\left[\frac{\pmb{\nabla}\psi}{\psi}\right]\!\big(\vb{R}(t),t\big)\times\vu{s}.
\end{align}
In the present case, the spatial wave function \(\psi\), Eq.\ \eqref{psit}, is a function of \(r\) only; therefore, 
\begin{align}\label{grad}
    \frac{\pmb{\nabla}\psi}{\psi} =\psi^{-1}\kern0.1em \frac{\mathrm{d}\psi}{\mathrm{d}\kern0.1em  r}\kern0.1em \vu{e}_{r}= \left[\frac{\pi}{a}\cot(\frac{\pi\kern0.1em  r}{a})-\frac{1}{r}\right]\kern-0.1em\kern0.1em \vu{e}_{r},\qquad r<a,
\end{align}
with \(\smash{\kern0.1em \vu{e}_{r}=r^{-1}\vb{r}}\). Since this quantity is real-valued, the first term on the right-hand side of \eqref{guide2} vanishes. Consequently, only the second nonzero term, i.e., the spin or Gordon term, determines the velocity of the electron within the box. Incorporating \eqref{grad} in Eq.\ \eqref{guide2}, we obtain
\begin{equation}\label{explicit}
    \dot{\vb{R}}(t)=\frac{\hbar}{m R(t)}\left[\frac{\pi}{a}\cot(\frac{\pi R(t)}{a})-\frac{1}{R(t)}\right]\vb{R}(t)\times\vu{s},
\end{equation}
where \(\smash{R(t)=\norm{\vb{R}(t)}}\) is the distance of the electron from the centre of the box at time \(t\). (The overdot indicates a derivative taken w.r.t.\ \(t\).) Equation \eqref{explicit} holds for \(\smash{R(t)<a}\); the electron's velocity is undefined for \(\smash{R(t)\ge a}\), i.e., outside and on the surface of the box (where its wave function is identically zero).

However, not surprisingly, \emph{no} solution of \eqref{explicit} starting within the volume of the box allows the particle to escape it. To establish this, note that
\begin{align}
    \frac{\mathrm{d}}{\mathrm{d}\kern0.1em  t}\kern0.1em  R^2(t) = 2\,\vb{R}(t)\cdot\dot{\vb{R}}(t) \propto \vb{R}(t)\cdot\big(\vb{R}(t)\times\vu{s}\big) = 0,
\end{align}
i.e., \(R(t)\) is a constant of the motion with \(\smash{R(t)=R(0)}\). The electron's trajectory thus lies on the surface of a sphere of radius \(R(0)\), concentric with the spherical box. In particular, \(\smash{R(0)<a}\) implies that \(\smash{R(t)<a}\) for \emph{all} \(t\).

Furthermore, since the component of velocity along the spin-vector \(\vu{s}\) is zero, we have
\begin{align}\label{const2}
    \frac{\mathrm{d}}{\mathrm{d}\kern0.1em  t}\kern0.1em  \big(\vb{R}(t)\cdot\vu{s}\big) &= \dot{\vb{R}}(t)\cdot\vu{s}=0.
\end{align}
This implies that \(\smash{\vb{R}(t)\cdot\vu{s}=\text{const.}=\vb{R}(0)\cdot\vu{s}}\), or equivalently,
\begin{equation}\label{plane}
    \big(\vb{R}(t)-\vb{R}(0)\big)\cdot\vu{s}=0.
\end{equation}
That is, \(\vb{R}(t)\) lies in a plane \emph{perpendicular} to \(\vu{s}\) that passes through the point \(\vb{R}(0)\).\footnote{Recall that the equation of a plane in the point-normal form is given by \(\smash{(\vb{r}-\vb{r}_{\scalebox{0.7}{0}})\cdot \vu{n}=0}\).} It can thus be inferred that the electron trajectory is part of a circle (formed by the intersection of this plane with the spherical surface of radius \(R(0)\)) whose centre is situated at a distance \(|\vb{R}(0)\cdot\vu{s}|\) from the centre of the box.

In fact, the motion is \emph{periodic}. To demonstrate this we solve Eq.\ \eqref{explicit} with initial condition \(\smash{\vb{R}(0)=\vb{R}_0}\), assuming \(\smash{R_0<a}\). Since we have established that \(\smash{R(t)=R(0)=R_0}\), Eq.\ \eqref{explicit} can be written as
\begin{equation}\label{simple}
    \dot{\vb{R}}(t) = -\,\omega_0~\vb{R}(t)\times\vu{s},
\end{equation}
where
\begin{equation}\label{angfreq}
    \omega_0 = \frac{\hbar}{mR_0^2}\left[1-\frac{\pi R_0}{a}\cot(\kern-0.1em\frac{\pi R_0}{a}\kern-0.1em)\right]\!,
\end{equation}
is a positive\footnote{\(\smash{x\cot x<1}\) for \(\smash{0<x<\pi}\).}, trajectory-specific constant, determined by the initial condition. (Note that Eq.\ \eqref{simple} is the familiar Bloch equation for the precession of a magnetic moment \(\vb{M}(t)\) in a uniform magnetic field \(\vb{B}\), with Larmor frequency \(\omega_0\).) Differentiating \eqref{simple} w.r.t.\ \(t\), we have
\begin{align}\label{2dot}
    \ddot{\vb{R}}(t) = -\,\omega_0\,\dot{\vb{R}}(t)\times\vu{s}\overset{\eqref{simple}}{=}\omega_0^2\,\big(\vb{R}(t)\times\vu{s}\big)\times\vu{s}=\omega_0^2\,\Big[\big(\vb{R}(t)\cdot \vu{s}\big)\,\vu{s}\,-\,\norm{\vu{s}}^2\,\vb{R}(t)\Big],
\end{align}
applying the triple vector cross product identity \(\vb{a}\times(\vb{b}\times\vb{c})=(\vb{a}\cdot\vb{c})\,\vb{b}-(\vb{a}\cdot\vb{b})\,\vb{c}\). Since \(\smash{\norm{\vu{s}}=1}\) and \(\vb{R}(t)\cdot\vu{s}\) is a constant of the motion, Eq.\ \eqref{2dot} reduces to
\begin{equation}
    \ddot{\vb{R}}(t) + \omega_0^2\,\vb{R}(t) = \omega_0^2\,\big(\vb{R}_0\cdot\vu{s}\big)\,\vu{s},
\end{equation}
the (classical) equation of motion for a forced simple harmonic oscillator. Its general solution is the familiar one:
\begin{equation}\label{general}
   \vb{R}(t)=\big(\vb{R}_0\cdot\vu{s}\big)\,\vu{s}+\cos(\omega_0t)\kern0.1em \vb{J} + \sin(\omega_0t)\kern0.1em \vb{K}, 
\end{equation}
where \(\vb{J}\) and \(\vb{K}\) are arbitrary constant vectors. Evidently, the motion is periodic with period \(2\kern0.1em \pi/\omega_0\), characteristic of an individual trajectory.

The variation with radial distance of the angular velocity \(\omega_0\), Eq.\ \eqref{angfreq}, reflects the profile of the wave function.\footnote{\label{omega}Electron orbits for the hydrogen ground-state wave function are similar in all respects to the present ones, except that \(\smash{\smash{\omega_{\scalebox{0.7}{0}}=\hbar/(ma_{\scalebox{0.7}{0}}R_{\scalebox{0.7}{0}})}}\) \cite[Eq.\ (7)]{Colijn}, where \(a_{\scalebox{0.7}{0}}\) is the Bohr radius. Those for an isotropic harmonic potential of trapping frequency \(\omega\) have \(\smash{\omega_{\scalebox{0.7}{0}}=\omega}\), independent of \(R_{\scalebox{0.7}{0}}\).} Electrons orbiting closer to the center of the box \(\smash{(R_0\ll a)}\) have an angular velocity
\begin{equation}
    \omega_0\sim \frac{\pi^2}{3}\kern0.1em \frac{\hbar}{ma^2}\,+\,\mathcal{O}\!\left(\kern-0.1em\frac{R_0}{a}\kern-0.1em\right)^{\!2},
\end{equation}
whereas for those moving close to the boundary of the box,
\begin{equation}\label{astronomical}
  \omega_0\sim \frac{\hbar}{ma(a-R_0)},  
\end{equation}
as \(R_0\) approaches \(a\). The latter suggests that the electron is moving at an astronomically high speed near the box's edge! Such a feature is not surprising in a nonrelativistic treatment. However, it is absent in the relativistic account presented below, where the dBB velocity \(\vb{v}^\Psi\) does not exceed \(c\) (the speed of light in vacuum).

\begin{figure}[!ht]
    \centering
    \includegraphics[width=7cm]{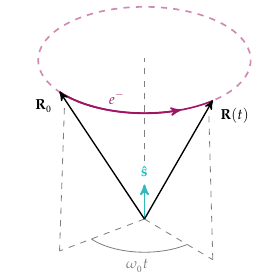}
    \caption{A de Broglie-Bohm electron trajectory for the spin-polarized ground-state wave function of a spherical box.}
    \label{fig1}
\end{figure}

To complete the trajectory calculation, note that \(\smash{\vb{R}(0)=\vb{R}_0\overset{\eqref{general}}{\Rightarrow}\vb{J}=\vb{R}_0-(\vb{R}_0\cdot\vu{s})\,\vu{s}}\). In order to determine \(\vb{K}\), differentiate \eqref{general} w.r.t.\ \(t\) and use the initial velocity condition \(\dot{\vb{R}}(0)\overset{\eqref{simple}}{=}-\,\omega_0\,\vb{R}_0\times\vu{s}\). This yields \(\smash{\vb{K}=-\,\vb{R}_0\times\vu{s}}\). The explicit solution for the Bohmian trajectory is thus given by
\begin{align}\label{rodrigues}
    \vb{R}(t)= \cos(\omega_0t)\,\vb{R}_0\,+\,\sin(\omega_0t)\,\vu{s}\times\vb{R}_0\,+\,\big[1-\cos(\omega_0t)\big]\,\big(\vb{R}_0\cdot\vu{s}\big)\,\vu{s},
\end{align}
which is reminiscent of Rodrigues' rotation formula. Geometrically, this means that the position of the particle at time \(t\) is obtained by rotating its initial position vector \(\vb{R}_0\) by an angle \(\omega_0t\) about the spin-vector-axis counterclockwise, as illustrated in Fig.\ \ref{fig1}. 
\section{Electron-in-a-box: Relativistic treatment}\label{RT}
In what follows, we present a brief relativistic account of the electron-in-a-box problem. For this, Dirac's Hamiltonian
\begin{equation}
    \mathcal{H} = -\,i\hbar\kern0.1em  c\,\bm{\alpha}\cdot\pmb{\nabla} + \beta\kern0.1em  mc^2 + V(r)
\end{equation}
replaces the Schr\"odinger-Pauli Hamiltonian \eqref{Ham} of the nonrelativistic theory, and the dBB velocity field is given by \cite{BohmT}, \cite[Ch.\ XVI]{dBBook}, \cite{Holland}
\begin{equation}\label{BD}
    \vb{v}_{\text{R}}^\Psi(\vb{r},t) = c\frac{\Psi^\dagger\bm{\alpha}\Psi}{\Psi^\dagger\Psi},
\end{equation}
where \(\Psi(\vb{r},t)\) is a four-spinor wave function. In these equations,
\begin{equation}\label{alphamat}
    \bm{\alpha}=\begin{pmatrix}\mathbbold{0} &\bm{\sigma}\\\bm{\sigma} &\mathbbold{0}\end{pmatrix} 
\end{equation}
is the \(\smash{4\times4}\) Dirac \(\alpha\)-vector matrix containing in the off-diagonal blocks the Pauli-spin-vector-matrix introduced earlier, \(\beta =\text{diag}(1,1,-1,-1)\), and \(\mathbbold{0}\) is the \(\smash{2\times2}\) zero or null matrix. This velocity field reproduces \eqref{velo} in the nonrelativistic limit. In fact, \(\vb{v}_{\text{R}}^{\Psi}\) can be rewritten to closely resemble \eqref{velo} (featuring spatial derivatives of the wave function) through Gordon's decomposition identity \cite[Sec.\ 8.1]{GreinerWaveEquations}.

That \(\smash{\norm*{\vb{v}_{\text{R}}^\Psi(\vb{r},t)}\le c}\) regardless of \(\Psi\) is a consequence of a so-called Fierz identity \cite{Fierznice}, \cite[Sec.\ 4]{Kofink}:
\begin{equation}\label{Firz}
   (\Psi^\dagger\Psi)^2-\norm*{\Psi^\dagger\bm{\alpha}\Psi}^2 = \mu^2+\nu^2,
\end{equation}
where \(\smash{\mu=\Psi^\dagger\beta\Psi}\), and \(\smash{\nu=\Psi^\dagger(\alpha_x\kern0.1em \alpha_y\kern0.1em \alpha_z)\beta\Psi}\) are \emph{real-valued} quantities. Clearly, the left-hand side of \(\eqref{Firz}\) is \(\geq 0\), establishing our claim.\footnote{Actually, the very form of the velocity field \eqref{BD} forces \(\norm*{\vb{v}_R^\Psi}\) to be \emph{bounded} for any \(\Psi\). This is a consequence of a theorem in linear algebra \cite[Thm.\ 4.2.2., p.\ 234]{RRThm} which states that for any nonzero \(\smash{\Xi\in\mathbb{C}^n}\), and \(n\times n\)-Hermitian matrix \(A\), the ratio \((\Xi^\dagger A \Xi)/(\Xi^\dagger \Xi)\) (the so-called Rayleigh quotient) is real-valued and \(\le \lambda_{\max}\)---the largest eigenvalue of \(A\). In light of this result and observing that \(\smash{\lambda_{\max}=1}\) for \(\alpha_x\), \(\alpha_y\), or \(\alpha_z\) (since each \(\alpha\) matrix squares to the identity matrix), it follows that every component of \eqref{BD} is \(\le c\), hence via the triangle inequality, \(\smash{\norm*{\vb{v}_R^\Psi}\le \sqrt{3}\, c}\). However, to establish the tighter bound \(\norm*{\vb{v}_R^\Psi}\le c\), additional input about the \(\alpha\) matrices is needed.}

Next, we address the electron in a spherical box problem. To begin, the relativistic ground-state wave function needs to be found. It turns out that we can resurrect the nonrelativistic ground-state wave function to obtain its relativistic counterpart (see, App.\ \ref{DiracSol} for details), with the final result
\begin{equation}\label{GSWF}
    \Psi(\vb{r},t) = e^{-\,itE_{\text{R}}/\hbar}\begin{pmatrix}\displaystyle\psi(\vb{r})\chi\\[2pt]\displaystyle-\,\frac{i\hbar\kern0.1em  c}{E_{\text{R}}+mc^2}\,\big(\bm{\sigma}\cdot\pmb{\nabla}\big)\kern0.1em \psi(\vb{r})\chi\end{pmatrix},
\end{equation}
where \(\psi(\vb{r})\) is the nonrelativistic spatial wave function \eqref{psit}, \(\chi\) is an arbitrary, constant two-spinor, and
\begin{equation}\label{ER}
    E_{\text{R}} = mc^2\sqrt{1+\frac{2E}{mc^2}},
\end{equation}
with \(E=\eqref{energy}\). (\(\smash{E_{\text{R}}\approx mc^2 + E}\) for \(\smash{E\ll mc^2}\).) Note that \eqref{GSWF} is not space-spin separated as its nonrelativistic counterpart \eqref{gs}. In what follows, we set
\begin{equation}\label{gamma}
    \frac{\hbar\kern0.1em  c}{E_{\text{R}}+mc^2}=\gamma
\end{equation}
for brevity (\(\gamma\) has the physical dimensions of a length).

Since \(\psi\) is real and \(\pmb{\nabla}\psi=\mathrm{d}\psi/\mathrm{d}\kern0.1em  r\, \kern0.1em \vu{e}_{r}\), we obtain
\begin{align}
    \Psi^\dagger\Psi &= \psi^2~\big(\chi^\dagger\chi\big)\,+\,\gamma^2\kern-0.1em\left(\frac{\mathrm{d}\psi}{\mathrm{d}\kern0.1em  r}\right)^{\!2}\!\big(\kern0.1em \vu{e}_{r}\cdot\bm{\sigma}\chi\big)^\dagger \big(\kern0.1em \vu{e}_{r}\cdot\bm{\sigma}\chi\big)\nonumber\\[3pt]
    &=\left(\psi^2\,+\,\gamma^2\left(\frac{\mathrm{d}\psi}{\mathrm{d}\kern0.1em  r}\right)^{\!2}\right)(\chi^\dagger\chi),
\end{align}
using \(\smash{\bm{\sigma}=\bm{\sigma}^\dagger}\) and the identity \eqref{identity}. Furthermore, following some simplifications,
\begin{equation}
    \Psi^\dagger\bm{\alpha}\kern0.1em \Psi = -\,i\gamma \left(\!\psi\kern0.1em \frac{\mathrm{d}\psi}{\mathrm{d}\kern0.1em  r}\right)\chi^\dagger\Big[\bm{\sigma}\kern0.1em \big(\kern0.1em \vu{e}_{r}\cdot\bm{\sigma}\big)\,-\,\big(\kern0.1em \vu{e}_{r}\cdot\bm{\sigma}\big)\kern0.1em \bm{\sigma}\Big]\kern0.1em \chi.
\end{equation}
The terms enclosed in square brackets evaluate to \(2\kern0.1em  i\,\kern0.1em \vu{e}_{r}\times\bm{\sigma}\). Thus, in terms of the unit spin-vector \eqref{spinvec} introduced earlier, we have
\begin{equation}\label{veloD}
\vb{v}_{\text{R}}^\Psi(\vb{r},t) = c\, F\!\left(\frac{\gamma}{\psi}\frac{\mathrm{d}\psi}{\mathrm{d}\kern0.1em  r}\right)\kern0.1em \vu{e}_{r}\times\vu{s},
\end{equation}
where \[F(\xi)=\frac{2\kern0.1em \xi}{1+\xi^2}.\]

Even without substituting the known expression for \(\psi\), it follows quite explicitly from \eqref{veloD} that \(\smash{\norm*{\vb{v}_{\text{R}}^\Psi}\le c}\), given \(\smash{\norm{\kern0.1em \vu{e}_{r}\times\vu{s}}\le 1}\), and \(\smash{|F(\xi)|\le 1}\) for any real \(\xi\). And, in the nonrelativistic limit \(\smash{c\to\infty}\),\footnote{More appropriately, when the nondimensionalised velocity \(\smash{(m\kern0.1em  a/\hbar)\,c\to\infty}\).} equivalently \(\smash{\gamma\to0}\), such that \(\smash{\gamma c\overset{\eqref{gamma}~}{\longrightarrow}\hbar/(2\kern0.1em  m)}\), the right-hand side of \eqref{veloD} obligingly reproduces the nonrelativistic velocity field, Eqs.\ (\ref{grad}-\ref{explicit}). The \emph{absolute necessity} of the spin-term of Eq.\ \eqref{velo} cannot be overemphasized here. Without it, the expected nonrelativistic limits would fail to follow.

It is a fortunate coincidence that the relativistic guiding equation obtained by incorporating \eqref{veloD} in \eqref{guide} has precisely the same form as its nonrelativistic counterpart, Eq.\ \eqref{explicit}. This implies that the relativistic electron orbits are circular as well, given by the closed-form expression \eqref{rodrigues}, except that \(\omega_0\) must be replaced by
\begin{equation}
    \omega_{0\text{R}} = \frac{c}{R_0}\kern0.1em  F\!\left(\frac{m R_0}{\hbar}\kern0.1em  \gamma\kern0.1em  \omega_0\right)\!,
\end{equation}
where \(\omega_0\) is the \(R_0\)-dependent nonrelativistic angular velocity of Eq. \eqref{angfreq}. Given that \(F\) is a bounded function with linear behaviour around zero, \(\omega_{0\text{R}}\) is bounded for all \(\smash{0\le R_0\le a}\). In particular,
\begin{equation}
    \omega_{0\text{R}} \sim \frac{2\kern0.1em  c}{\gamma} \left(1-\frac{R_0}{a}\right) \,+\, \mathcal{O}(a-R_0)^3,
\end{equation}
as \(\smash{R_0\to a^-}\), thus taming the divergence encountered in \eqref{astronomical}.

\begin{figure}[!ht]
    \centering
    \includegraphics[width=9cm]{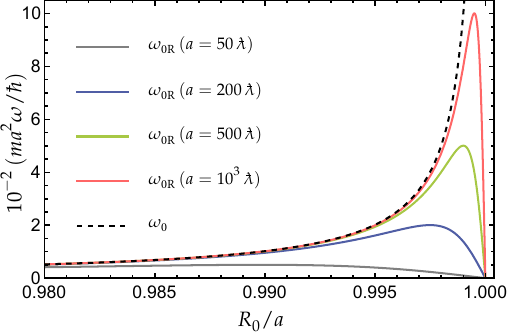}
    \caption{Nondimensionalized relativistic (\(\omega_{0\text{R}}\)) and nonrelativistic (\(\omega_0\)) angular velocities vs.\ \(R_0\), the electron's distance from the centre of the box. Here, \(a\) and \(\,\smash{{\mkern0.75mu\mathchar '26\mkern -9.75mu\lambda}=\hbar/(mc)}\) denote the box radius and the reduced Compton wavelength of the electron, respectively. Notable differences between \(\omega_0\) and \(\omega_{0\text{R}}\) only appear near the edge of the box for \(\smash{R_0>0.9\kern0.1em  a}\). Note that \(\omega_{0\text{R}}\,(\omega_0)\) vanishes (blows up) as \(\smash{R_0\to a^-}\).}
    \label{fig2}
\end{figure}

The relativistic and nonrelativistic angular velocities are compared in Fig.\ \ref{fig2} for select values of \(a\). The relativistic orbital velocity increases monotonically with increasing distance from the box's centre, reaching a maximum value (characteristic of the box radius \(a\)), after which it begins to decrease and eventually vanishes asymptotically as \(\smash{R_0}\) approaches \(a\). In other words, the electrons slow down to rest at the box's boundary. Numerical calculations show that \(\omega_{0\text{R}}\) is appreciably different from \(\omega_{0}\) only for \(a-R_0\lesssim \hbar/(mc)\approx3.86\times10^{-13}\kern0.1em \)m (the reduced Compton wavelength). Therefore, FAPP, the ground-state electron orbits are adequately described by the nonrelativistic theory.
\section{Electron motion post-deconfinement}\label{deconfinement}
Left to itself, the trapped electron will continue orbiting indefinitely, leaving little more of interest to say about this system. To evoke a more dramatic response, one could imagine opening the box suddenly, say, at time \(t_0\), and subsequently investigating the wave function of the particle, and the evolution of its motion, due to such a ``shock.''  Specifically, this could aid in the analysis of an experiment designed to catch the emerging electron on, e.g., a microchannel plate.

The quantities that are typically reported in this kind of experiment are either the electron's impact position, its arrival time, or both. These measurements are ubiquitous in particle physics and serve as the basis for measuring momentum, energy, etc. The measurement of momentum will be treated in detail in Sec.\ \ref{momentumTOF}. In what follows, we present a nonrelativistic discussion of the electron trajectories (using the dBB equations of Sec.\ \ref{NRT}) post-deconfinement.

In the absence of the spherical box potential \eqref{sphericalwell}, the wave function evolves freely\footnote{We are neglecting, for brevity, the electron's free fall due to gravity in the problem.} with Hamiltonian \(\mathcal{H}=-\,\hbar^2/(2\kern0.1em  m)\nabla^2\kern0.1em  \mathbbold{1}\) (cf.\ Eqs.\ \eqref{Ham} and \eqref{identity}), thus continuing to remain space-spin separated, as \eqref{gs}, for \(\smash{t>t_0}\). Over time, the electron is transported to distances \(\smash{\gg a}\) as the wave function spreads isotropically, filling the entire space. The wave function for any \(t\) can be expressed as follows:
\begin{equation}
    \Psi(\vb{r},t) = \begin{cases}
        \,\eqref{gs}, &t\le t_0\\ \,\psi_{\scaleto{>}{3.5pt}}(\vb{r},t)\chi, & t>t_0
    \end{cases},
\end{equation}
where \(\psi_{\scaleto{>}{3.5pt}}\) solves the free-particle Schr\"odinger equation 
\begin{equation}\label{freesch}
    i\hbar\kern0.1em \frac{\partial\psi_{\scaleto{>}{3.5pt}}}{\partial t}=-\,\frac{\hbar^2}{2\kern0.1em  m}\nabla^2\psi_{\scaleto{>}{3.5pt}}. 
\end{equation}
In order for \(\Psi\) and \(\vb{v}^\Psi\) to be continuous at \(t=t_0\), we require
\begin{equation}\label{initial}
    \psi_{\scaleto{>}{3.5pt}}\big(\vb{r},t_0^+\big) = e^{-\,it_{\scalebox{0.5}{0}} E/\hbar}\,\psi(\vb{r}),
\end{equation}
where \(\psi(\vb{r})\) is the spatial-part of the ground state wave function, Eq.\ \eqref{psit}.

\begin{figure}[!ht]
     \includegraphics[width=15.5cm]{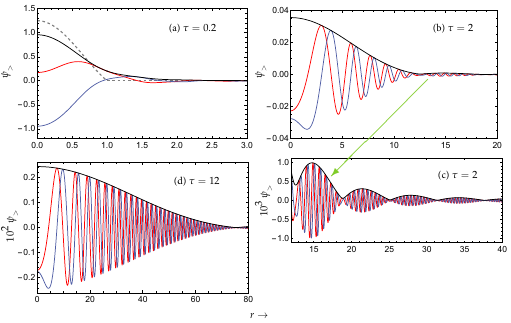}
    \caption{Snapshots of the spatial wave function \(\psi_{\scaleto{>}{3.5pt}}\): \(\Re[\psi_{\scaleto{>}{3.5pt}}]\) (red), \(\Im[\psi_{\scaleto{>}{3.5pt}}]\) (blue), and \(|\psi_{\scaleto{>}{3.5pt}}|\) (black) for  \(\smash{a=1}\) at select instants of \(\tau\). The dashed curve in panel (a) denotes the spatial ground state wave function.}
    \label{fig3}
\end{figure}

The solution of Eq.\ \eqref{freesch} with initial condition \eqref{initial} can be found in the literature; see, e.g., \cite{DITsphere,Godoy0,Godoy}. Nonetheless, for completeness, we derive it in App.\ \ref{postswitch} by methodically reducing the three-dimensional calculation to a one-dimensional one that can be solved with ease, obtaining
\begin{align}\label{psig}
    \psi_{\scaleto{>}{3.5pt}}(\vb{r},t) = \frac{N_0}{r}\kern0.1em \Big[M(r+a,k,\tau)-M(r-a,k,\tau)-\kern0.1em  M(a-r,k,\tau)+M(-\kern0.1em  a-r,k,\tau)\Big].
\end{align}
Here, 
\begin{equation}\label{constants}
    N_0=\frac{i\kern0.1em  e^{-\,iEt_{{\scalebox{0.5}{0}}}/\hbar}}{2\kern0.1em \sqrt{2\pi a}},\qquad k=\frac{\pi}{a},\qquad \tau=\frac{\hbar}{m}\kern0.1em  (t-t_0),
\end{equation}
and \(M(r,k,t)\) is Moshinsky's function, which inevitably appears in problems involving matter waves suddenly released from bounded domains \cite{Mfunction,QTransients}. Note that the electron wave function ceases to be \emph{stationary} post-deconfinement, i.e., it is no longer separated in \(\vb{r}\) and \(t\). A few snapshots of \(\psi_{\scaleto{>}{3.5pt}}\) are plotted in Fig.\ \ref{fig3}.

Besides the general dispersive spreading of the wave function, with its amplitude gradually diminishing over time at every \(r\), \(\psi_{\scaleto{>}{3.5pt}}\) presents a train of tiny ripples, or wavelets, close to the boundary of the spherical box. With time, these wavelets move farther away, nucleating into distinct, self-similar wave packets. This wave phenomenon, dubbed \emph{diffraction in time}---due to its ``close mathematical resemblance with the intensity of light in the Fresnel diffraction by a straight edge'' \cite{Moshinsky}, arises in response to a sudden change in the boundary conditions of the wave function at some surface (that of the spherical box, \(\smash{r=a}\), in the present case).\footnote{Note that \(\psi_{\scaleto{>}{3.5pt}}\) becomes nonzero at \emph{all} distances from the spherical trap as soon as the trap is removed. A relativistic treatment of the problem expectedly modifies this feature, requiring the wave function to strictly vanish for any \(\smash{r>a+ c\kern0.1em  (t-t_0)}\), at time \(t\) \cite[Sec.\ IV]{Moshinsky}, \cite{GodoyVilla}.}

Some of these aspects of the wave function are easier understood considering the approximation
\begin{equation}\label{psigapprox}
    \psi_{\scaleto{>}{3.5pt}}(\vb{r},t) \approx 2\kern0.1em  N_0\,\sqrt{\frac{2\kern0.1em \pi}{i\tau}}\,\frac{a}{r}\,\frac{\sin(ar/\tau)}{(ar/\tau)^2-\pi^2}\,e^{i(r^2 \kern0.1em +\kern0.1em  a^2)/(2\tau)},
\end{equation}
which is derived in App.\ \ref{postswitch} assuming \(\smash{r\gg a}\) and \(\smash{r\gg \pi \tau/a}\). Interestingly, though, with the exception of very small values of \(\tau\), it turns out to be a very good approximation to \eqref{psig} for practically all \(r\) and \(\tau\).

Now, turning to the Bohmian trajectories, the simplified guiding equation \eqref{guide2} (applicable for spin-polarized wave functions) continues to be valid for \(\smash{t>t_0}\) except that \(\smash{\psi\to\psi_{\scaleto{>}{3.5pt}}}\). As \(\psi_{\scaleto{>}{3.5pt}}\) is spherically symmetric, we write
\begin{equation}
    \frac{\pmb{\nabla}\psi_{\scaleto{>}{3.5pt}}}{\psi_{\scaleto{>}{3.5pt}}} = \Big[\mathcal{R}(r,t)+i\,\mathcal{I}(r,t)\Big]\kern0.1em \kern0.1em \vu{e}_{r},
\end{equation}
where \(\mathcal{R}\) (\(\mathcal{I}\)) is the real (imaginary) part of \(\psi_{\scaleto{>}{3.5pt}}^{-1}\kern0.1em  \mathrm{d}\psi_{\scaleto{>}{3.5pt}}/\mathrm{d}r\). This allows expressing the equation of motion for \(\smash{t>t_0}\) in the form
\begin{equation}\label{guide3}
    \dot{\vb{R}}(t) = \frac{\hbar}{m R(t)}\,\Big( \mathcal{I}\big(R(t),t\big)\kern0.1em \vb{R}(t)\,+\,\mathcal{R}\big(R(t),t\big)\kern0.1em \vb{R}(t)\times\vu{s}\Big),
\end{equation}
where \(\vu{s}\) is the unit spin-vector defined in Eq.\ \eqref{spinvec}. (For \(t\le t_0\), Eq.\ \eqref{explicit} applies.) Note that the first term within parentheses (originating from the convective velocity term) is now \emph{nonzero}. Also, the velocity field is now explicitly time-dependent, making it much harder to determine Bohmian trajectories.

For convenience, let us use a spherical-polar coordinate system where the polar axis is pointing along \(\vu{s}\). The particle position and velocity at time \(t\) are given by (cf.\ \cite[Ch.\ 2]{sphericalkinematics} or \cite[Sec.\ 3.5]{Symon71})
\begin{equation}\label{velocitypol}
\vb{R}=R\kern0.1em \kern0.1em \vu{e}_{r},\qquad\dot{\vb{R}}=\dot{R}\kern0.1em \kern0.1em \vu{e}_{r}\,+\,R\,\dot{\Theta}\kern0.1em \vu{e}_{\theta}\,+\,R\kern0.1em \sin\Theta\,\dot{\Phi}\kern0.1em \kern0.1em \vu{e}_{\phi},
\end{equation}
respectively, where \(R(t)\), \(\Theta(t)\) and \(\Phi(t)\) denote the spherical coordinates of the particle at time \(t\). We also have \(\smash{\vu{s} = \cos\theta\kern0.1em \kern0.1em \vu{e}_{r}-\sin\theta\kern0.1em \vu{e}_{\theta}}\), hence \(\smash{\kern0.1em \vu{e}_{r}\times\vu{s} = -\sin\theta\kern0.1em \big(\kern0.1em \vu{e}_{r}\times\vu{e}_{\theta}\big)= -\sin\theta\kern0.1em \kern0.1em \vu{e}_{\phi}}\). Incorporating this into \eqref{guide3} and comparing the result with \(\dot{\vb{R}}\), Eq.\ \eqref{velocitypol}, yields the component equations
\begin{align}\label{3EOM}
    \dot{R}(t) = \frac{\hbar}{m}\, \mathcal{I}\big(R(t),t\big),\qquad 
    \dot{\Theta}(t) = 0,\qquad
    \dot{\Phi}(t) =-\,\frac{\hbar}{m R(t)}\, \mathcal{R}\big(R(t),t\big).
\end{align}

The second equation has the obvious solution \(\smash{\Theta(t)=\Theta(t_0)}\), the value of \(\Theta\) at time \(t_0\). In fact, \(\Theta\) being the angle of inclination of the electron's position vector \(\vb{R}\) to the spin-vector \(\vu{s}\) (see Fig.\ \ref{fig1}) is a constant of the motion for \(\smash{t<t_0}\) as well.

\begin{figure}[!ht]
    \centering
    \includegraphics[width=\columnwidth]{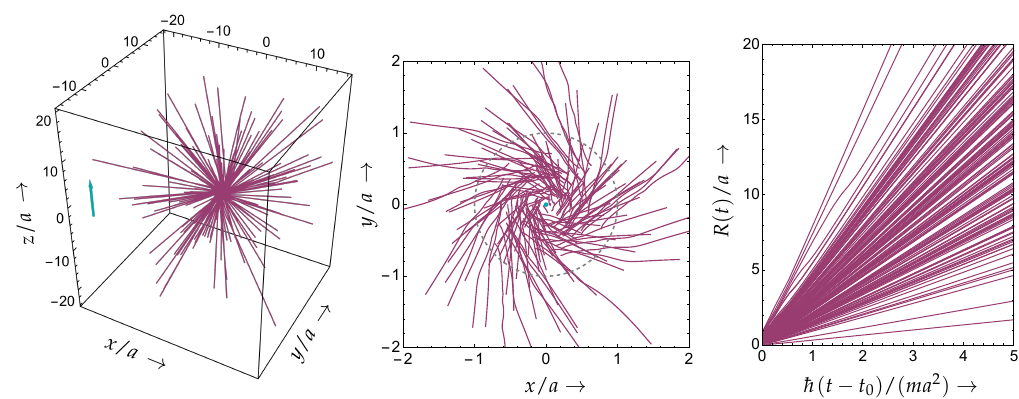}
    \caption{Left panel: A collection of Bohmian trajectories emanating from the spherical box, obtained by numerically integrating Eqs.\ \eqref{3EOM} for a random sample of initial positions. The spin-vector \(\vu{s}\) is parallel to the arrow depicted in the figure. Middle panel: Bird's-eye view of the trajectories in the near-field with \(\vu{s}\) pointing out of the page. The dashed circle marks the boundary of the spherical trap that was released at time \(t_0\). Right panel: Displacement vs.\ time graphs of the same paths demonstrating how quickly the electron's radial velocity approaches a constant.}
    \label{fig4}
\end{figure}

To this point our considerations have been exact. Unfortunately, the remaining two equations in \eqref{3EOM} cannot be solved analytically. We may, however, glean the qualitative behaviour of the Bohmian trajectories in the far field (i.e., \(\smash{r\gg a}\)), which is relevant for our subsequent discussion of momentum measurements, below. For this, we utilize \eqref{psigapprox} to obtain without further approximations \(\mathcal{I}(r,t)\approx r/\tau\), in turn, \(\smash{\dot{R} \approx R(t)/(t-t_0)}\). The solution is 
\begin{equation}\label{asymptotic}
    R(t) \approx v_\infty (t-t_0),
\end{equation}
where \(v_\infty\) is a constant of integration. The escaping electron thus acquires a \emph{constant radial velocity} at large separations from the spherical box; see Fig.\ \ref{fig4} (right panel). The constant \(v_\infty\) is trajectory-specific and varies with the exact initial position of the electron.\footnote{For an initial Gaussian wave function of width \(\sigma_0\), the guiding equations can be solved exactly, yielding \(v_\infty = \hbar \kern0.1em  R_0/(2\kern0.1em  m\sigma_0^2)\) \cite[Eq.\ (8.7)]{HollandPhilippidis}. That is, the further the electron starts from the centre of the wave packet, the faster it escapes.}

The remaining equation for \(\dot{\Phi}\) can be similarly analysed using \eqref{psigapprox}. One finds in view of \eqref{asymptotic} that \(\smash{\Phi(t)\to \text{const.}}\) asymptotically. Individual Bohmian trajectories therefore become approximately straight lines of constant \(\Theta\) and \(\Phi\) in the far field, as shown in Fig.\ \ref{fig4} (left panel). In other words, they become \emph{asymptotically Newtonian with a well-defined constant velocity}, which is a prominent feature of Bohmian trajectories in scattering problems \cite{SarahRomer,DDGZ96,DDGZ}, \cite[Ch.\ 16]{DurrTeufel}.

Note: The spin velocity contribution is more pronounced in the near-field (i.e., at distances \(\smash{\lesssim a}\)), where the trajectories tend to circulate \(\vu{s}\), as shown in Fig.\ \ref{fig4} (middle panel). However, it decreases with distance as \(\mathcal{R}(r,t)\) becomes subleading to \(\mathcal{I}(r,t)\) in the far field. As a result, it is the convective velocity primarily governing the motion of the unbound electron post-deconfinement, whereas the spin velocity alone is responsible for the motion of the bound electron (Sec.\ \ref{NRT}).\footnote{Although this pattern is somewhat paradigmatic (see also \cite{HollandPhilippidis}), the spin velocity can dominate the electron's motion over very long distances in some special cases, such as when a spin-polarized electron is propagating within a waveguide and its motion is bounded in all but one direction \cite{Exotic}.}
\section{Momentum spectroscopy: General considerations}\label{momentumTOF}
For a particle described by the wave function \(\Psi(\vb{r})\), the standard/textbook quantum formalism predicts that the result of a momentum measurement is random with distribution \(|\tilde{\Psi}(\vb{p})|^2\) (or \(\tilde{\Psi}^\dagger\tilde{\Psi}\) for spinor-valued \(\Psi\)), where
\begin{equation}\label{FT}
    \tilde{\Psi}(\vb{p}) = \big(2\kern0.1em \pi\hbar\big)^{-\,3/2}\int_{\mathbb{R}^3}d^3r~\Psi(\vb{r})\exp(\!-\,\frac{i}{\hbar}\,\vb{p}\cdot\vb{r})
\end{equation}
is the momentum-space wave function. It is easily calculated for the ground state of the spherical box (Eq.\ \eqref{gs}):
\begin{equation}\label{sinc}
    \tilde{\Psi}(\vb{p}) = \left(\frac{a}{\hbar}\right)^{3/2}\!\frac{\sin q}{q\kern0.1em  \big(\pi^2-q^2\big)}\,\chi,\qquad q=\frac{a}{\hbar}\kern0.1em  \norm{\vb{p}}.
\end{equation}
However, the question of what precisely constitutes a ``momentum measurement'', i.e., what experiments could validate the above prediction, is often passed over in silence. 


On the other hand, a plethora of experimental literature exists on \emph{recovering} the momentum distribution (or the closely related Wigner distribution) from directly observable quantities, like the impact position or arrival time of the particle \cite{COLTRIMS,RIMS,Pfau,Hediffraction}. But since \emph{reconstruction is always theory-dependent} we must, in the end, turn to theory to relate the directly measured quantities to the desired ones. To discuss this point further, we take a closer look at the widely used ``time-of-flight momentum spectroscopy'' technique.


The key step of this method involves a measurement of the transit time or time-of-flight (ToF) of a particle prepared in the wave function of interest at a starting time \(t_0\) and allowed to move freely. When this particle is registered by a remote detector at a (random) time \(t_1\), its ToF
\begin{equation}\label{ToF}
    t_f=t_1-t_0
\end{equation}
is used to define the particle's momentum, or, rather, its magnitude \(p\), as follows:
\begin{equation}\label{defn}
    p = \frac{m L}{t_f}.
\end{equation}
Here, \(L\) is the known distance between the source and the detector (typically much larger than the width of the prepared wave packet). After repeating the experiment numerous times and accumulating the flight times, an empirical distribution of \(t_f\), and in turn \(p\) given by Eq. \eqref{defn}, is obtained. The latter is expected to reproduce the quantum mechanical distribution
\begin{equation}\label{marginal}
    \Lambda_{\text{QM}}(p) =\int_{\mathbb{R}^3}d^3p^\prime~\delta\big(\norm*{\vb{p}^\prime}-p\big)\,\tilde{\Psi}^\dagger\big(\vb{p}^\prime\big)\tilde{\Psi}\big(\vb{p}^\prime\big),
\end{equation}
where \(\delta\) is Dirac's delta function. In mathematical terms, if \(\Pi_{\text{exp}}(t_f)\) is the measured ToF distribution, one expects
\begin{equation}\label{hope}
   \Lambda_{\text{QM}}(p)= \frac{m L}{p^2}\,\Pi_{\text{exp}}\!\left(\!\frac{m L}{p}\!\right)\!.
\end{equation}

At the core of this scheme lies Eq.\ \eqref{defn}, \emph{which translates the measured \(t_f\) to the reported \(p\).} It is motivated under the assumption that free Newtonian motion prevails, enabling the particle to approach the detector at a constant speed \(p/m\); \(p\) being the magnitude of the particle's momentum. Heisenberg, for instance, makes just such an assumption in \cite[p.\ 32]{Heisenberg}:
\begin{quote}
    ``The momentum of the [bound] electron can most readily be measured by suddenly rendering the interaction of the electron with the nucleus and neighbouring electrons negligible. \emph{It will then execute a straight-line motion} and its momentum can be measured in the manner already explained.'' (Emphasis added.)
\end{quote}
Considering how passionately Heisenberg contended that since motion and particle trajectories are not disclosed to us outside of measurement, hence speaking of them has essentially no meaning or advantage \cite{Aristarhov}, it may seem incredibly circular that he refers to motion in the very conception of momentum measurements. Indeed, according to \cite{ValentiniMWBook}, Einstein cautioned Heisenberg that his treatment of observation was unduly laden with the outdated theory of classical mechanics, and this would eventually get him into ``hot water''.

However, in all fairness to Heisenberg and other quantum physicists who availed themselves of Newtonian presuppositions in the early stages of theory development, one could argue \`a la \cite[pp.\ 503-504]{ValentiniMWBook}, paraphrasing Einstein, that while
\begin{quote}
    ``experiment is theory-laden, and correct measurement procedures must be laden with the correct theory, [...] when new experimental phenomena are discovered---phenomena that require the formulation of a new theory---in practice the old theory is at first assumed to provide a reliable guide to interpreting the observations [...] Note that this is a practical necessity, for the new theory has yet to be formulated. However---and here is the crucial point---once the new theory \emph{has} been formulated, one ought to be careful to use the new theory to design and interpret measurements, and not continue to rely on the old theory to do so. For one may well find that consistency is obtained only when the new laws are found \emph{and applied to the process of observation.}''
\end{quote}
But today, nearly a century later, many quantum physicists boldly reject the existence of particle trajectories,\footnote{Echoing, e.g., \cite[p.\ 2]{LandauLifshitz}: ``In quantum mechanics there is no such concept as the path of a particle.''} even though trajectory-based considerations are essential to numerous experiments designed to verify the predictions of quantum mechanics, including the ToF momentum spectroscopy method under consideration.

An alternative way to justify the ToF method without using classical trajectories would start with the quantum mechanical prediction \(\Pi_{\text{QM}}(t_f)\) for the directly measured ToF distribution, and deduce that Eq. \eqref{defn} is simply a change of variables that relates it to the momentum distribution via Eq.\ \eqref{hope} (with \(\smash{\Pi_{\text{exp}}\to \Pi_{\text{QM}}}\)). However, this is easier said than done because calculating the quantum mechanical ToF distribution has been a long-standing issue, absent a recognized arrival-time observable, with various proposed solutions \cite{MUGA1,MSP,gaugeinv}.\footnote{It has even been suggested that the ``time of arrival cannot be precisely defined and measured in quantum mechanics'' \cite{stupid} or ``that wave mechanics cannot accommodate an exact and ideal arrival-time concept'' \cite{Allcock3}. Should these suggestions be taken seriously, Eq. \eqref{defn} becomes vacuous.} Here, we will not go into a case-by-case examination of the existing suggestions to determine which ones validate or contradict the experimental method in question. 

But, against this background, it is clear that the dBB quantum theory involving particle trajectories, and not dependent on quantum observables to describe experiments, would be more appropriate for evaluating this experimental technique. In the following section, we offer a dBB analysis of the ToF momentum spectroscopy method by applying it to the electron-in-a-box example.
\section{ToF momentum spectroscopy: A dBB account}\label{anapplication}
For mathematical convenience, we will assume the ToF detector has a spherical surface with radius \(\smash{L\gg a}\) that surrounds the spherical box. Once the box is abruptly removed at time \(t_0\), as shown in Sec.\ \ref{deconfinement}, the trapped electron veers from its otherwise circular orbit and begins moving along a spiral curve that eventually follows a rectilinear path. An individual electron, therefore, contacts the detector (resulting in a detection event) at a definite time \(t_1\) at which
\begin{equation}\label{implicit}
    R(t_1)=L,
\end{equation}
where \(R(t)\) denotes the electron's distance from the centre of the box at time \(t\). Equations \eqref{implicit} and \eqref{ToF} implicitly define the ToF of the electron as a function of its initial position \(\vb{R}_0\) within the box (which determines \(R(t)\) via the guiding Eqs.\ \eqref{explicit} and \eqref{3EOM}), and of the known parameters \(L\) and \(t_0\).

While we are only able to determine this \(t_f\) numerically (see below), for \(\smash{L\gg a}\) one expects 
\begin{equation}\label{cool}
    t_f \approx \frac{L}{v_\infty},
\end{equation}
where \(v_\infty\) is the asymptotic Bohmian velocity of the electron (cf.\ Sec.\ \ref{deconfinement}); keeping in mind that the brief period spent in the near-field close to the trapping region contributes negligibly to the total ToF. It thus follows from \eqref{cool} that the \(p\) determined via \eqref{defn} is simply the asymptotic dBB momentum of the electron acquired in flight, post deconfinement, \emph{viz.}, \(m \kern0.1em  v_\infty\).\footnote{Although, the momentum of a Bohmian particle defined along Newtonian lines as \(\smash{m\kern0.1em  \dot{\vb{R}}}\) is generally not conserved even in the absence of external potentials.} Just this insight, free from implicit Newtonian assumptions, provides a much-needed {\emph {physical foundation}} for Eq.\ \eqref{defn} in quantum mechanical contexts.

Although a single electron can be dependably prepared in the ground state of a trap at the start of each experimental run, controlling its position is practically impossible; for this reason, the measured arrival time \(t_1\) varies randomly from one experimental run to the next. The statistical distribution of \(t_1\), and hence \(t_f\), may be calculated thanks to the \emph{quantum equilibrium hypothesis} (QEH), which postulates that the random initial positions of a Bohmian particle follow Born's statistical rule, see \cite{DGZ,TravisBornrule,CALLENDERBornrule,SColin}. In particular, the electron positions within the box are distributed according to the \(\Psi^\dagger\Psi\) distribution, where \(\Psi\) is the ground state wave function \eqref{gs}.\footnote{In order for \(\Psi^\dagger\Psi\) to be a legitimate probability distribution, \(\chi^\dagger\chi=1\).}

\begin{figure}[!ht]
    \centering
    \includegraphics[width=\columnwidth]{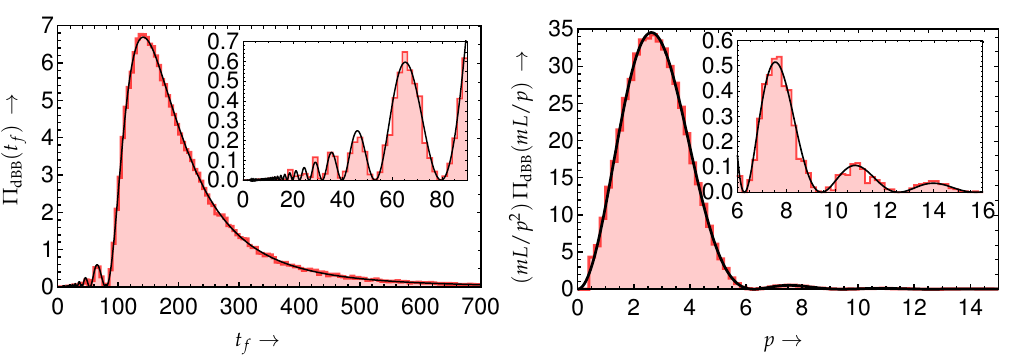}
    \caption{In this figure masses, lengths, and times are expressed in units of the electron mass \(m\), the box radius \(a\) and \(ma^2/\hbar\), respectively. Left panel: ToF Histogram made from \(\approx 1.25\times 10^5\) randomly sampled Bohmian trajectories emanating from the spherical box and arriving on a spherical detector of radius \(\smash{L=500}\). Right panel: histogram of \(p\) reconstructed from these arrival times using the formula \eqref{defn}. The solid line depicts \(\Lambda_{\text{QM}}(p)\) defined in \eqref{lambdaQM}.}
    \label{fig5}
\end{figure}

Leveraging the QEH, Fig.\ \ref{fig5} (left panel) presents a ToF histogram generated from \(\smash{\approx 1.25\times 10^5}\) Bohmian trajectories, the flight times of which were acquired by numerically integrating the guiding equations. The main lobe of the histogram is preceded by many smaller, secondary lobes (seen magnified in the inset), which are caused by a small fraction of electrons that get rapidly transported by the diffraction-in-time wavelets discussed in Sec.\ \ref{deconfinement} (see also \cite{Mousavi,Exotic}). ToF experiments involving cold neutrons and atoms have also reported the presence of such secondary lobes \cite{DIT,DIT1}. 

As in the real experiments, we convert the computed Bohmian flight times into momenta using Eq.\ \eqref{defn}, and the result is the momentum histogram displayed in Fig.\ \ref{fig5} (right panel). The quantal prediction \eqref{marginal}, given by
\begin{equation}\label{lambdaQM}
    \Lambda_{\text{QM}}(p) \overset{\eqref{sinc}}{=} \frac{4\kern0.1em  \pi a}{\hbar}~\frac{\sin^2(pa/\hbar)}{\big[\pi^2-(pa/\hbar)^2\big]^2}
\end{equation}
is fully in agreement with it. Since the dBB theory does not even utilize momentum-space wave functions in stating its dynamical equations, this agreement is quite remarkable indeed.

We offer here some theoretical calculations that corroborate the numerical results, turning first to a consideration of the electron's arrival time \(t_1\). The distribution of this quantity, \(\Pi_{\text{dBB}}(t_1)\), is usually very difficult to evaluate, but in the current scenario, where the Bohmian trajectories intercept the detection surface \emph{at most} once, it is directly expressible in terms of the wave function as explained, e.g., in \cite[Sec.\ 6]{gaugeinv}. We find for this distribution, the exact result 
\begin{align}\label{gmdBB}
    \Pi_{\text{dBB}}(t_1) = \frac{\hbar}{m}\,\big(4\kern0.1em  \pi L^2\big)\kern0.1em   \Im\!\left[\psi_{\scaleto{>}{3.5pt}}^*\frac{d\psi_{\scaleto{>}{3.5pt}}}{d r}\right]\!(L,t_1).
\end{align}
Letting \(\smash{t_1\overset{\eqref{ToF}}{\to}t_f+t_0}\), \eqref{gmdBB} becomes the ToF distribution depicted by the black curve in Fig.\ \ref{fig5} (left panel). Recalling our approximation for \(\psi_{\scaleto{>}{3.5pt}}\), Eq.\ \eqref{psigapprox}, one can see immediately that the obtained ToF distribution would reproduce \eqref{lambdaQM} exactly when incorporated into Eq.\ \eqref{hope} in place of \(\Pi_{\text{exp}}(t_f)\). 

Our explanation of the ToF momentum spectroscopy technique for the particle-in-a-box scenario is now complete. The theoretical treatment proceeds along similar lines in most cases. That being said, it is important to be aware of exceptional situations where the technique would {\emph {not}} work. For instance, in situations where the initial wave functions are not well-localized, \cite{WardNico}, or when there is backflow, \cite{DD}, or where more than one entangled particle is involved, e.g., \cite{AliJavadDDslit}, the Bohmian trajectories are not necessarily asymptotically Newtonian, hence the applicability of \eqref{defn} is questionable. 
\section{Conclusion}\label{conclusion}  
The dBB theory views spin as an aspect of the movement of quantum particles dictated by guiding equations that incorporate spinor-valued wave functions. To illustrate this spin-aware motion of fermions, the first part of this chapter provided a comprehensive, step-by-step examination of the motion of a single electron confined in a spherical box using both relativistic and nonrelativistic versions of the theory. Next, we investigated the reaction of the trapped electron to a sudden removal of the confining potential.

The chapter's second part focused on how the dBB theory is typically applied to actual experiments. For concreteness, we investigated the widely used ToF momentum spectroscopy method, which is grounded in trajectory-based assumptions. It was noted that standard quantum mechanics, which prohibits discussing particle trajectories and is beset with the long-standing ``arrival-time problem,'' makes such assumptions extremely difficult to defend. On the other hand, the dBB treatment of this experimental method is transparent and produces the anticipated results in a principled way, offering a solid theoretical basis for the heuristics that experimenters are employing to interpret the measurements. A few potential situations in which the heuristics might be insufficient were also touched upon.
\section*{Acknowledgements}
The initial segment of the chapter originated from conversations with Prof.\ Jean Bricmont, for whose unwavering support and several thought-provoking exchanges I am very grateful. Many thanks to James M.\ Wilkes for his insightful editing suggestions that led to a significant improvement of the text. Thanks also to the anonymous reviewer for pointing out a few errors and for helpful suggestions. Finally, I would like to express my gratitude to Andrea Oldofredi for inviting me to contribute to this special volume. 
\appendix
\section{Relativistic ground-state wave function}\label{DiracSol}
The electron's ground-state wave function assumes the form \(\smash{\Psi(\vb{r},t)=\Psi(\vb{r})\,e^{-\,itE_{\text{R}}/\hbar}}\), where \(\Psi(\vb{r})\) is a four-spinor satisfying Dirac's time-independent equation
\begin{equation}\label{TIDE}
    \big(\kern-0.1em-i\hbar\kern0.1em  c\,\bm{\alpha}\cdot\pmb{\nabla}+mc^2\beta\big)\kern0.1em \Psi=E_{\text{R}}\Psi
\end{equation}
in the region \(\smash{r<a}\). Here, \(\smash{\bm{\alpha}=\eqref{alphamat}}\), and \(\smash{\beta=\text{diag}(\mathbbold{1},-\mathbbold{1})}\). Denoting the upper and lower two-spinor components of \(\Psi\) by \(\Psi_\pm\), Eq.\ \eqref{TIDE} yields the component equations
\begin{equation}\label{pm}
    -i\hbar\kern0.1em  c\,(\bm{\sigma}\cdot\pmb{\nabla})\kern0.1em \Psi_\mp = \big(E_{\text{R}}\,\mp\, mc^2\big)\kern0.1em \Psi_\pm.
\end{equation}
Eliminating \(\Psi_-\), we obtain
\begin{equation}\label{SPD}
    -(\bm{\sigma}\cdot\pmb{\nabla})^2\kern0.1em \Psi_+=\frac{E_{\text{R}}^2-m^2c^4}{\hbar^2c^2}\kern0.1em  \Psi_+.
\end{equation}
The 2-spinor \(\Psi_+\) becomes the wave function of the electron in the nonrelativistic limit. Noting that \(\smash{(\bm{\sigma}\cdot\pmb{\nabla})^2=\nabla^2\mathbbold{1}}\), cf.\ Eq.\ \eqref{identity}, we find that Eq.\ \eqref{SPD} reduces to the usual time-independent Schr\"odinger-Pauli equation 
\begin{equation}\label{PSE}
    -\frac{\hbar^2}{2\kern0.1em  m}\nabla^2\kern0.1em  \Psi_+ = E\kern0.1em \Psi_+,
\end{equation}
upon setting 
\begin{equation}\label{EV}
    \frac{E_{\text{R}}^2-m^2c^4}{\hbar^2c^2} = \frac{2\kern0.1em  m}{\hbar^2}\kern0.1em  E.
\end{equation}
The ground-state wave function solution of \eqref{PSE}, which corresponds to \(\smash{E=\eqref{energy}}\), has already been identified, \emph{viz.}, \(\smash{\Psi_+=\psi(\vb{r})\chi}\) with \(\psi(\vb{r})=\eqref{psit}\), \(\chi\) being a constant spinor. Equation \eqref{EV} then fixes \(E_{\text{R}}\) in terms of the nonrelativistic ground-state energy \(E\). Note that only \(\smash{E_{\text{R}}>0}\) fulfilling \eqref{EV} (Eq.\ \eqref{ER}) pertains to electron solutions. Finally, incorporating \(\Psi_+\) into Eq.\ \eqref{pm}, we obtain \(\Psi_-\), and hence the complete ground-state wave function \eqref{GSWF}. 
\section{Time-evolution post switching}\label{postswitch}
The free-particle Schr\"odinger equation \eqref{freesch} with initial condition \eqref{initial} can be solved as follows: Exploiting the spherical symmetry of the problem, express the solution as
\begin{equation}\label{ansatz}
    \psi_{\scaleto{>}{3.5pt}}(\vb{r},t) = r^{-1}\,\varphi\!\left(r,\frac{\hbar}{m}\,(t-t_0)\right)\!.
\end{equation}
We require 
\begin{equation}\label{initialH}
    \varphi(r,0) \overset{\eqref{initial}}{=} r\kern0.1em  \psi_{\scaleto{>}{3.5pt}}(\vb{r},t_0)=\frac{H(a-r)}{\sqrt{2\kern0.1em \pi a}}\sin(\frac{\pi\kern0.1em  r}{a}) e^{-\,it_{\scalebox{0.5}{0}}E/\hbar}.
\end{equation}
(\(H(\cdot)\) denotes Heaviside's step function.) In order for Eq.\ \eqref{ansatz} to provide a solution \(\psi_{\scaleto{>}{3.5pt}}\) of Eq.\ \eqref{freesch}, the function \(\varphi\) must satisfy the one-dimensional Schr\"odinger equation
\begin{equation}\label{1Dsch}
    i\frac{\partial\varphi}{\partial t}=-\,\frac{1}{2}\kern0.1em \frac{\partial^2\varphi}{\partial r^2}.
\end{equation}
Furthermore, given our \emph{ansatz} \eqref{ansatz}, \(\varphi\) must vanish at \(\smash{r=0}\) in order for \(\psi_{\scaleto{>}{3.5pt}}\) to be regular at the origin.

We are then essentially seeking the wave function of a free particle moving on the half-line \(\smash{(r\ge0)}\) subject to an impenetrable potential barrier at \(\smash{r=0}\), and specified initial condition at time zero. The solution can be expressed in the form
\begin{equation}\label{propK}
    \varphi(r,t) = \int_0^{\infty}\!\!dr^\prime~\varphi\big(r^\prime,0\big)\kern0.1em  K\big(r,r^\prime,t\big),
\end{equation}
where 
\begin{equation}
    K\big(r,r^\prime,t\big) =-\, \sqrt{\frac{2\kern0.1em  i}{\pi t}}\sin(\kern-0.1em\frac{rr^\prime}{t}\kern-0.1em)\exp[\frac{i}{2\kern0.1em  t}\kern0.1em \big(r^2+r^{\prime\kern0.1em 2}\big)]
\end{equation}
\cite{propagator,halfline}. Letting \(\smash{u=\pi/a}\) and \(\smash{v=r/t}\), the trigonometric identity \(2\kern0.1em  \sin(u r^\prime)\sin(v r^\prime)=\cos[(u-v)\kern0.1em  r^\prime]\,-\,\cos[(u+v)\kern0.1em  r^\prime],\) allows rewriting \eqref{propK} as 
\begin{equation}\label{step2}
    \varphi(r,t) = N_0\kern0.1em \Big[\eta(r,t)\,-\,\eta(-\kern0.1em  r,t)\Big],
\end{equation}
where \(\smash{N_0=ie^{-\,iEt_{{\scalebox{0.5}{0}}}/\hbar}/\big(2\kern0.1em \sqrt{2\kern0.1em \pi a}\,\big)}\) and
\begin{equation*}
    \eta(r,t)=i\kern0.1em \sqrt{\frac{2\kern0.1em  i}{\pi t}}\int_0^a\!dr^\prime\exp[\frac{i}{2\kern0.1em  t}\kern0.1em \big(r^2+r^{\prime\kern0.1em 2}\big)]\cos(\kern-0.1em\frac{\pi r^\prime}{a}-\frac{rr^\prime}{t}\kern-0.1em)
\end{equation*}
---a Gaussian integral expressible in terms of the familiar error (or complementary error) function. However, it is convenient to express the final result in terms of the Moshinsky function \(M(r,k,t)\) \cite{Mfunction}, which is a judicious repackaging of the complementary error function:
\begin{equation}\label{MM}
    M(r,k,t) = \frac{1}{2}\,\text{erfc}\kern-0.1em\left(\!\frac{r-kt}{\sqrt{2it}}\!\right) e^{ikr\,-\,it k^2/2}.
\end{equation}
(the function \(M(r,k,t)\) itself satisfies Eq.\ \eqref{1Dsch}, reducing to the ``truncated plane wave'' \(H(-\kern0.1em  r)\kern0.1em  e^{ikr}\) in the limit \(\smash{t\to 0^+}\).)\, In any case, we have
\begin{equation}
    \eta(r,t) = M\!\left(r-a,\frac{\pi}{a},t\right)-M\!\left(r+a,\frac{\pi}{a},t\right)\!,
\end{equation}
which via Eqs.\ \eqref{ansatz} and \eqref{step2} completely defines the time-evolved wave function \(\psi_{\scaleto{>}{3.5pt}}(\vb{r},t)\), Eq.\ \eqref{psig}.

For  \(|r-kt|\gg 1\), the Moshinsky function has the following approximation:
\begin{equation}
    M(r,k,t) \approx \sqrt{\frac{it}{2\kern0.1em \pi}}\,\frac{e^{i r^2/(2t)}}{r-kt}\, + \,H(kt-r)\kern0.1em  e^{ikr\,-\,it k^2/2},
\end{equation}
from which it follows, for \(\smash{r\gg a}\) and \(\smash{r\gg\pi t/a}\), the approximations
\begin{equation}
    \eta(\pm\kern0.1em  r,t) \approx  \sqrt{\frac{2\kern0.1em  t}{i\pi}}\,\frac{\sin(ra/t)}{r \mp \pi t/a}\,e^{i(r^2+a^2)/(2t)},
\end{equation}
in turn, using Eqs.\ \eqref{step2} and \eqref{ansatz}, the result \eqref{psigapprox}.
\bibliography{sample}
\end{document}